\documentclass[12pt]{article}
\usepackage{float, epsfig}

\begin{document}
	
\title{
Markov properties of high frequency exchange rate data
}

\author{
Ch. Renner, J. Peinke\\
Fachbereich 8 Physik, Universit\"at Oldenburg, D-26111 Oldenburg\\
R. Friedrich,\\
Institut f\"ur Theoretische Physik,\\
Universit\"at Stuttgart, D-70550 Stuttgart, \\
}
\maketitle

\begin{abstract}
	
	We present a stochastic analysis of a data set consisiting of $10^6$
	quotes of the US Doller - German Mark exchange rate. Evidence is
	given that the price changes $x(\tau)$ upon different delay times
	$\tau$ can be described as a Markov process evolving in $\tau$. Thus,
	the $\tau$-dependence of the probability density function (pdf)
	$p(x,\tau)$ on the delay time $\tau$ can be described by a
	Fokker-Planck equation, a gerneralized diffusion equation for
	$p(x,\tau)$. This equation is completely determined by two
	coefficients $D_{1}(x,\tau)$ and $D_{2}(x,\tau)$ (drift- and diffusion
	coefficient, respectively). We demonstrate how these coefficients can
	be estimated directly from the data without using any assumptions or
	models for the underlying stochastic process. Furthermore, it is
	shown that the solutions of the resulting Fokker-Planck equation
	describe the empirical pdfs correctly, including the pronounced
	tails.\\

\end{abstract}
\noindent
keywords: econophysics, Markov processes, FX data

\section{Introduction}

Since L. Bachelier's pioneering work dating back to 1900
\cite{Bachelier}, the complex statistical properties of economic
systems have attracted the attention of many researchers and an
extensive literature has evolved for modeling fluctuations of
financial markets. Traditionally, fluctuations of financial assets
were viewed and modeled as random variables. Well known examples are
the ARCH-type models (see for example, \cite{Engel92, Bol86, Bol92})
and stochastic volatility (SV) models (\cite{Taylor94, Ghysels96,
Gallant}).

Since advances in computer technology have made high frequency data
available, many physicists have joined the field adapting methods from
statistical physics. One line of studies within econophysics focusses
on the statistical properties of financial time series such as stock
prices, stock market indices or currency exchange rates. Rather than
comparing the predictions of models with the various aspects of
emprical data (which is the traditional approach), physicists try to
extract information about the stochastic processes governing financial
markets from an analysis of emprical data. An overview of recent
developments in this field can be found in references \cite{Stanley}
, \cite{Stanley2000} and \cite{Voit}. In spite of all the efforts spent, some of
the most basic questions concerning the statistics of financial assets
remain unsolved. In particular, the mechanism leading to the
fat-tailed (leptokurtic) probability distributions of fluctuations on 
small time scales is still unknown. Compared to a Gaussian, the pdfs 
of those fluctuations express an unexpected high probability for large 
fluctuations (see figure \ref{TheoExp1}). Quantifying the risks of 
such large fluctuations is of utmost importance for the risk 
management and the pricing of options.

Fluctuations of financial time series $y(t)$ are usually measured by 
means of returns, log-returns or, equivalently, by price increments. 
Here, we consider the statistics of the price increment $x(\tau)$ over 
a certain time scale $\tau$, which is defined as:
\begin{eqnarray}
    x(\tau) \; = \; y(t+\tau) - y(t). \label{incdef}
\end{eqnarray}
We suppressed dependence of the price increment $x(\tau)$ on the time $t$ 
since we assume the underlying stochastic process to be stationary. 

In a recent letter \cite{PRL}, we demonstrated that the mathematics of
stochastic processes is a useful tool for empirical investigations of
the time scale dependence of the proabability density function (pdf)
$p(x,\tau)$ of the price increment $x$ on the time scale
$\tau$\footnote{The correct notation for the probability density
function of the stochastic variable $x(\tau)$ is $p(x(\tau))$.  We
write $p(x,\tau)$ in order to indicate that the pdf of $x(\tau)$
depends on the parameter $\tau$.}. It was shown how the equations
governing the underlying stochastic process can be extracted directly
from the empirical data, provided that the price increment $x$ obeys a
Markov process. In particular, it is possible to derive a partial
differential equation, the Fokker-Planck equation, which describes the
evolution of the pdf $p(x,\tau)$ in the scale variable $\tau$.  Hence,
the mathematics of Markov processes yields a complete description of
the stochastic process underlying the evolution of the pdfs from
Gaussian distributions at large scales $\tau$ to the leptokurtic pdfs
at small scales.

Here, we extend our previous analysis presented in \cite{PRL}
considerably. We show how the existence of a Markov process can be
checked empirically and how the Fokker-Planck equation can be
calculated directly from the data. Furthermore, we show how
multiplicative and additive noise sources interact leading to the
leptokurtic distributions and volatility clustering of the exchange
rates. Finally, our results are interpreted in the context of
currently discussed features of financial time series. Since our
method is not based on any models, we gain new insights into the
mechanisms governing the statistics of economic systmes. In
particular, we find new evidence for cascading processes in financial
markets.

The paper is organized as follows. Section \ref{theory} is devoted to
a brief summary of the most important notions and theorems on Markov
processes and their application to the analysis of empirical data,
while section \ref{analysis} contains the main results of our
analysis. We shall consider in detail a data set containing $10^6$
samples of the DM / US-Dollar exchange rate from the one-year period
october '92 to september '93.  An interpretation of our results is
finally given in section \ref{discussion}.

\section{Mathematics of Markov Processes} \label{theory}

The following section briefly summarizes the notions and theorems
which will be of importance for our statistical analysis. For further
details on Markov processes we refer the reader to the references 
\cite{Risken} and \cite{Hae}.

Fundamental quantities related to Markov processes are conditional
probability density functions. Given the joint probability density
$p(x_{1},\tau_{1};x_{2},\tau_{2})$ for finding the price increments $x_{1}
\, (\, := \, x(\tau_{1}) \,)$ at a scale $\tau_{1}$ and $x_{2}$ at a
scale $\tau_{2}$ with $\tau_{1}<\tau_{2}$, the conditional pdf
$p(x_{1},\tau_{1}|x_{2},\tau_{2})$ is defined as:
\begin{eqnarray}
        p(x_{1},\tau_{1}|x_{2},\tau_{2}) \;=\;
        \frac{p(x_{1},\tau_{1};x_{2},\tau_{2})}{p(x_{2},\tau_{2})} \; . \label{cpdfdef1}
\end{eqnarray}
$p(x_{1},\tau_{1}|x_{2},\tau_{2})$ denotes the conditional probability
density for the price increment $x_{1}$ at a scale $\tau_{1}$ given an increment
$x_{2}$ at a scale $\tau_{2}$. It should be noted that the small scale
$\tau_{1}$ lies within the larger scale $\tau_{2}$ (see equation
(\ref{incdef}): the increments $x_{1}$ and $x_{2}$ are calculated for
the same time $t$).

Higher order conditional probability densities can be defined in
an analogous way:
\begin{eqnarray}
        p(x_{1},\tau_{1}|x_{2},\tau_{2}; ... ;x_{N},\tau_{N}) \; = \; \frac{p(x_{1},\tau_{1};
        x_{2},\tau_{2}; ... ;x_{N},\tau_{N})}{p(x_{2},\tau_{2}; ... ;x_{N},\tau_{N})} \;.
        \label{cpdfdef2}
\end{eqnarray}
Again, the smaller scales $\tau_{i}$ are nested inside the larger scales
$\tau_{i+1}$ (with the common reference point $t$).

The stochastic process in $\tau$ is a Markov process, if the conditional
probability densities fulfill the following relations:
\begin{eqnarray}
	p(x_{1},\tau_{1}|x_{2},\tau_{2};x_{3},\tau_{3};\ldots
	x_{N},\tau_{N}) \; & = & \;
	p(x_{1},\tau_{1}|x_{2},\tau_{2}) \;  \label{markovcondtheo}
\end{eqnarray}
with $\tau_{1} < \tau_{2} < \tau_{3} < \ldots< \tau_{N}$.  As a consequence of
(\ref{markovcondtheo}), each N-point probability density
$p(x_{1},\tau_{1};x_{2},\tau_{2};\ldots;x_{N},\tau_{N})$ can be determined as a
product of conditional probability density functions:
\begin{equation}\label{chain}
	p(x_1,\tau_1;...x_N,\tau_N)=p(x_1,\tau_1|x_2,\tau_2)....
	p(x_{N-1},\tau_{N-1}|x_N,\tau_N) 
	p(x_N,\tau_N) \; .
\end{equation}

Equation (\ref{chain}) indicates the importance of the conditional pdf
for Markov processes.  Knowledge of $p(x,\tau|x_{0},\tau_{0})$ (for
arbitrary scales $\tau$ and $\tau_{0}$ with $\tau<\tau_{0}$) is
sufficient to generate the entire statistics of the price increment
encoded in the N-point probability density
$p(x_{1},\tau_{1};x_{2},\tau_{2};x_{3},\tau_{3}; \ldots
;x_{N},\tau_{N}) $.

For Markov processes the conditional probability density fulfills a
master equation which can be put into the form of a Kramers-Moyal
expansion \footnote{Note that, in contrast to the usual definition as,
for example, given in \cite{Risken}, we multiplied both sides of the
Kramers-Moyal expansion by $\tau$ (see also equation (\ref{Mndef})). 
This is equivalent to the logarithmic length scale $-ln(\Delta
t / 40960s)$ used in \cite{PRL}. The negative sign of the left side
of equation (\ref{krammoyalexpand}) is due to the direction of the
cascade torward smaller scales $\tau$.}:
\begin{eqnarray}
- \, \tau \, \frac{\partial}{\partial \tau} \,
p(x,\tau|x_{0},\tau_{0}) \;=\;
\sum\limits_{k=1}^{\infty} \; \left( \, - \frac{\partial}{\partial
x} \,
\right)^k \, D_{k}(x,\tau) \,  p(x,\tau|x_{0},\tau_{0}). \label{krammoyalexpand}
\end{eqnarray}
The Kramers-Moyal coefficients $D_{k}(x,\tau)$ are defined as the limit
$\Delta \tau \rightarrow 0$ of the conditional moments $M_{k}(x,\tau,\Delta \tau)$:
\begin{eqnarray}
	D_{k}(x,\tau) &\,= \,& \lim_{\Delta \tau \rightarrow 0} \, M_{k}(x,\tau,\Delta
	\tau) \; , \label{dndef} \\
	M_{k}(x,\tau,\Delta \tau) &\,=\,& \frac{\tau}{k! \, \Delta \tau} \,
	\int\limits_{-\infty}^{+\infty} \, \left( \tilde{x} - x \right)^k \,
	p\left( \tilde{x}, \tau - \Delta \tau | x,\tau \right) \, d \tilde{x} .
\label{Mndef}
\end{eqnarray}
For a general stochastic process, all Kramers-Moyal coefficients are
different from zero. According to Pawula's theorem, however, the
Kramers-Moyal expansion stops after the second term, provided that the
fourth order coefficient $D_{4}(x,\tau)$ vanishes.  In that case, the
Kramers-Moyal expansion reduces to a Fokker-Planck equation (also
known as the backwards or second Kolmogorov equation):
\begin{equation}
        - \tau \frac{\partial}{\partial \tau} \, p(x,\tau|x_{0},\tau_{0}) \;=\;
        \left\{ \, - \frac{\partial}{\partial x} D_{1}(x,\tau) \, + \,
        \frac{\partial^2}{\partial x^2} D_{2}(x,\tau) \, \right\}
        p(x,\tau|x_{0},\tau_{0}) .  \label{foplacond}
\end{equation}

$D_{1}$ is denoted as drift term, $D_{2}$ as diffusion term. The
probability density $p(x,\tau)$ has to obey the same equation:
\begin{equation}
        - \, \tau \, \frac{\partial}{\partial \tau} \,
	p(x,\tau) \;=\;
        \left\{ \; - \, \frac{\partial}{\partial x} \, D_{1}(x,\tau) \; + \;
        \frac{\partial^2}{\partial x^2} \, D_{2}(x,\tau) \; \right\} \;
        p(x,\tau).  \label{foplauncond}
\end{equation}
The Fokker-Planck equation describes the probability density function
of a stochastic process generated by the Langevin equation (we use
It\^{o}'s definition)
\begin{equation} \label{Langevin}
    -\tau \frac{\partial }{\partial \tau}x(\tau)=D_1(x,\tau) +
    \sqrt{D_2(x,\tau)} f(\tau)
\end{equation}
where $f(\tau)$ is a Langevin force, i.e. $\delta$-correlated white noise
with a gaussian distribution. Here, the increment $x(\tau)$ at a fixed
time $t$ is generated by a stochastic process with respect to the
continuous variable $\tau$. This kind of process expresses nothing 
else but a hierarchical, cascade-like structure connecting price 
increments on different time scales \cite{JFM}.

In the case that the random force $f(\tau)$ ist not Gaussian
distributed, the Kramers-Moyal expansion does not reduce to the
Fokker-Planck equation. From the more general Kramers-Moyal expansion
(\ref{krammoyalexpand}), which is also valid for the probability
density $p(x,\tau)$, differential equations for the n-th order moments
can be derived. By multiplication with $x^n$ and integration with
respect to $x$ we obtain:
\begin{eqnarray}
        - \tau \frac{\partial}{\partial \tau} \left< x^n(\tau) \right> &\, =\,&
        \sum\limits_{k=1}^{\infty} \, \left( -1 \right)^k \,
        \int\limits_{-\infty}^{+\infty} \, x^n \, \left( \,
        \frac{\partial}{\partial x} \, \right)^k \, D_{k}(x,\tau) \, 
        p(x,\tau) dv
        \nonumber \\ &\,=\,& \sum\limits_{k=1}^{n} \, \frac{n!}{(n-k)!} \,
        \int\limits_{-\infty}^{+\infty} \, x^{n-k} \, D_{k}(x,\tau) \, 
        p(x,\tau) dv
        \; . \nonumber \\
		&\,=\,& \sum\limits_{k=1}^{n} \, \frac{n!}{(n-k!)} \, \left< \, 
		x^{n-k} \, D_{k}(x,\tau) \, \right> \; .
		\label{momentengleichung}
\end{eqnarray}

\section{Data Analysis} \label{analysis}

The hypothesis concerning the Markovian properties of exchange rate
data immediately fixes a framework for the analysis of the data. 
First, one has to give evidence of the Markovian properties according
to equation (\ref{markovcondtheo}). Secondly, the evolution of
conditional probability densities $p(x_1,\tau_1|x_2,\tau_2)$ has to be
specified on the basis of the stochastic evolution equation, eq. 
(\ref{krammoyalexpand}). To this end, we have to determine the
conditional moments $M_{k}(x,\tau,\Delta \tau)$ at different scales
$\tau$ for various values of $\Delta \tau$.

Practically, it is only possible to evaluate the lowest order
coefficients. Therefore, we shall restrict our analysis to the
coefficients of order one, two, and four. Approximating the limit
$\Delta \tau \rightarrow 0$, we obtain the Kramers-Moyal coefficients
$D_{k}(x,\tau)$. If the fourth order coefficient vanishes, the
evolution equation (\ref{krammoyalexpand}) takes the form of a
Fokker-Planck equation, according to Pawula's theorem. Since we
consider a finite data set, we are not able to prove rigorously 
(in a mathematical sense) that $D_{4}$ is zero but can only give 
hints for the validity of the assumption that the conditional 
probability density obeys a Fokker-Planck equation.

In order to verify this assumption and our results for the Kramers-Moyal 
coefficients, we compare the numerical solutions of the Fokker-Planck 
equations (\ref{foplacond}) and (\ref{foplauncond}) with the probability 
density functions determined directly from the data.

For the rest of the article, prices and price increments are given in
units of the standard deviation $\sigma$ of $y(t)$. For the data set under
consideration, $\sigma$ is $0.064 DM$.

\subsection{Data Processing} \label{processing}

\begin{figure}[ht]
  \begin{center}
    \epsfig{file=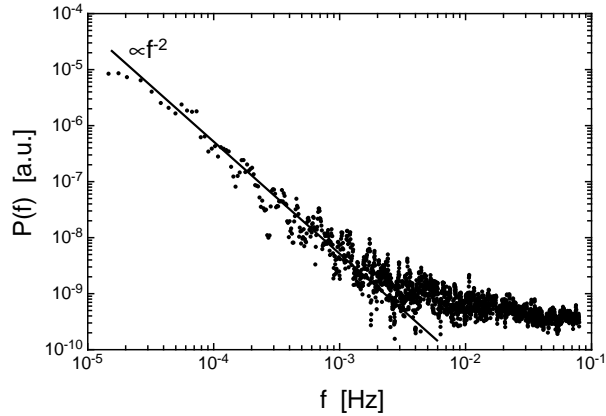, width=8.0cm}
  \end{center}
  \caption{Frequency spectrum of the exchange rate data $y(t)$. For 
  high frequencies the data are dominated by white noise.
 }
 \label{PowerSpectrum}
\end{figure}

From the frequency spectrum (figure \ref{PowerSpectrum}) of the 
exchange rate $y(t)$, it becomes evident that the data are dominated 
by white noise for high frequencies (above $\approx 10^{-2} Hz$), 
comparable to the noise affecting any physical measurement. The 
original signal $y(t)$ seems to be the sum of the underlying "real" 
signal $s(t)$ and some additive white noise $n(t)$:
\begin{equation}
    y(t) \;=\; s(t) + n(t).  \label{AddNoise}
\end{equation}
The additive noise acts on the increments in a similar way (see 
equation (\ref{incdef})). Note that this noise differs from the 
dynamical noise of the Langevin equation (\ref{Langevin}). $n(t)$ can be 
regarded as some randomness which is added seperately to the dynamical 
stochastic cascade process. In physics this kind of noise is known as 
measurement noise \cite{measurenoise}.

%
%

\begin{figure}[ht]
  \begin{center}
    \epsfig{file=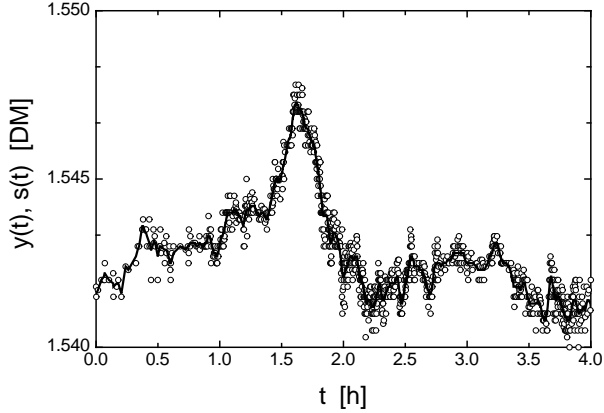, width=8.0cm}
  \end{center}
  \caption{Samples of the time series of the original exchange rate data $y(t)$ (cirles) 
  and the smoothed signal $s(t)$ (full line).
 }
 \label{Zeitserien}
\end{figure}

As will be seen in capter \ref{KMCoeffs}, the existence of such an 
additive white noise dominating the small scales causes problems in 
our analysis, especially in the limit $\Delta \tau \rightarrow 0$ 
which has to be performed according to equation (\ref{dndef}). 
In order to avoid these problems, we applied a low-pass filter to the 
data. Each value $y(t)$ was replaced by the weighted average of 
itself and its neighbouring data points, where the weighting function 
was chosen to be a gaussian centered at $t$ with a width of $44sec$.
Figure \ref{Zeitserien} shows the result $s(t)$ of this procedure in 
comparison to the original data $y(t)$.

%
%

According to equation (\ref{AddNoise}), we can extract the noise
$n(t)$ by subtracting the smoothed signal from the original
data $y(t)$. If the assumptions leading to equation (\ref{AddNoise})
are correct and the algorithm used to smooth the data is an
appropriate one, the extracted $n(t)$ should be white noise, i.e. it
should be $\delta$-correlated with zero mean. 

\begin{figure}[H]
  \begin{center}
    \epsfig{file=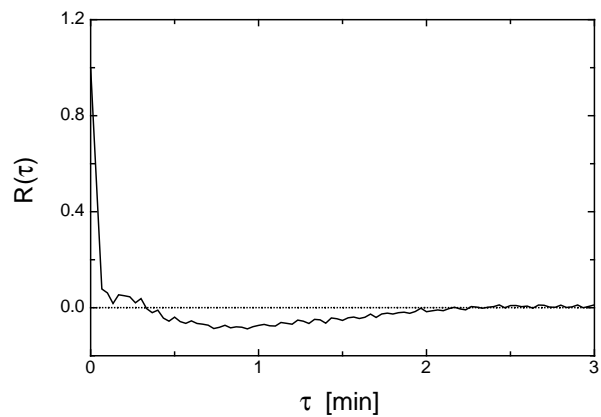, width=8.0cm}
  \end{center}
  \caption{The autocorrelation function of the extracted additive 
  noise $n(t)$.
 }
 \label{AutocorrNoise}
\end{figure}

The ratio of the mean value $<n(t)>$ to the standard deviation
$\sigma_{n}$ of $n(t)$ was found to be smaller than $8 \cdot 10^{-5}$
($\sigma_{n}=0.0003DM$). This result justifies the assumption of a
zero mean for $n(t)$.
The autocorrelation function $R_{n}(\tau)$ of $n(t)$ decreases from
$1$ to $0.08$ within two seconds (see figure \ref{AutocorrNoise}). 
Compared to the temporal resolution of the data (the smallest time step
between two subsequent data points is two seconds), $R(\tau)$ indeed
decreases rapidly. However, there appear to be small but
significantly non-zero values for time delays $\tau \leq 2min$, 
indicating that the approximation of the "real" signal $s(t)$ by the 
smoothed signal is insufficient.
We therefore restrict our analysis to time delays $\tau$ larger than a
certain elementary step $\tau_{min}$ which, from figure
\ref{AutocorrNoise}, we chose to be $4min$. In units of
$\tau_{min}$, $R_{n}(\tau)$ can be considered to be
$\delta$-correlated. 

%
%

\begin{figure}[ht]
  \begin{center}
    \epsfig{file=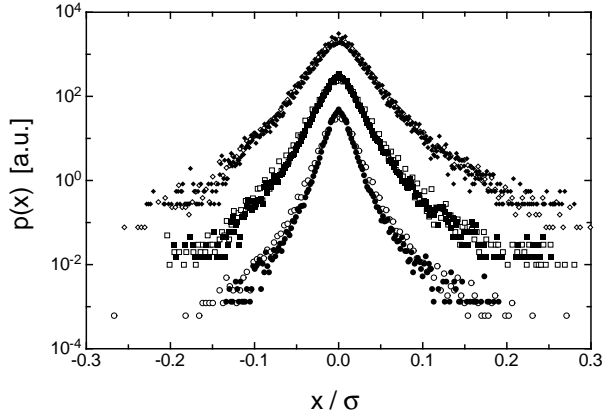, width=8.0cm}
  \end{center}
  \caption{Pdfs of price increments $x(\tau)$ for $\tau=4min, 8min$ and
  $15min$ (from bottom to top). The pdfs of $x$ were calculated from the
  smoothed data (full symbols) as well as from the original time series
  $y(t)$ (open symbols). Curves have been shifted in y-direction for
  clarity of presentation. Note that $\sigma$ ist not the respective 
  standard deviation of each individual pdf, but the standard 
  deviation of the whole data set $y(t)$.
 }
 \label{PdfVergleich}
\end{figure}

Figure \ref{PdfVergleich} finally compares the probability density 
functions of price increments $x(\tau)$ calculated from the original 
data $y(t)$ and the smoothed data $s(t)$. It becomes evident that the 
influence of the noise $n(t)$ is indeed restricted to small scales. For 
scales larger than $\tau=2\tau_{min}$ the pdfs are practically identical.

It may be worth to recall the result for fully developed turbulence,
where a similar stochastic analysis was performed for velocity
increments over space scales instead of price increments over time
scales. In the case of turbulence, the Markovian properties were
found to be valid only for (length-) scales and differences of scales
larger than an elementary step size $l_{mar}$ \cite{JFM}. But in
contrast to the elementary step in turbulence which is a physical
effect caused by the smoothing effects of viscous forces, the step
$\tau_{min}$ used here for the analysis of financial data is due to
the additional noise $n(t)$ as discussed above. We 
want to mention that it is not unusual to find such a "cut off" at 
small scales for stochastic processes. For Brownian motion, the mean 
free path of molecules defines an elementary step size for the 
diffusion process in an analogous way.

%
%

For the rest of the article, the price increments $x_{i}(\tau_{i})$ are 
(unless it is mentioned explicitly) calculated from the smoothed 
data $s(t)$.

\subsection{Markov Properties}

\begin{figure}[H]
  \begin{center}
    \epsfig{file=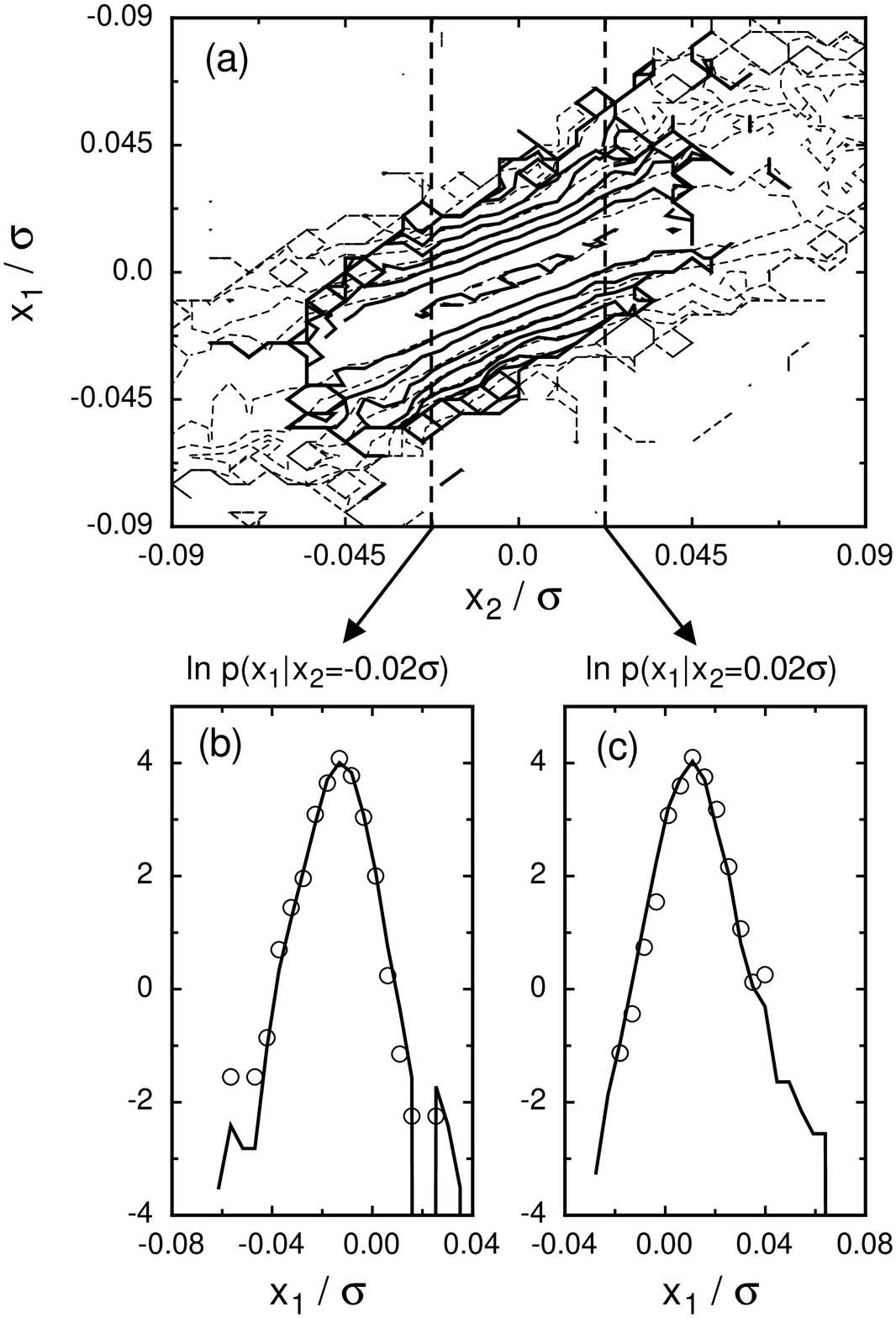, width=6.0cm}
  \end{center}
  \caption{(a): Contour plots of the conditional pdfs
  $p(x_{1},\tau_{1}|x_{2},\tau_{2})$ (dashed lines) and
  $p(x_{1},\tau_{1}|x_{2},\tau_{2};x_{3}=0,\tau_{3})$ (solid lines) for
  $\tau_{1}=\tau_{min}$, $\tau_{2}=2\tau_{min}$ and 
  $\tau_{3}=3\tau_{min}$. \newline (b) and (c): Cuts
  through the conditional pdfs for $x_{2}=\pm 0.02\sigma$. Open symbols:
  $p(x_{1},\tau_{1}|x_{2},\tau_{2};x_{3}=0,\tau_{3})$, solid lines:
  $p(x_{1},\tau_{1}|x_{2},\tau_{2})$.
 }
 \label{MarkovTestI}
\end{figure}

Next, we give evidence that the Markovian property
(\ref{markovcondtheo}) holds. Strictly speaking, the relationships
(\ref{markovcondtheo}) have to be verified for all positive values of
N as well as for each set of scales $\tau_{1},...,\tau_{N}$, a task,
which is evidently impossible. However, the data set considered here
consisting of $10^6$ samples allows to verify condition
(\ref{markovcondtheo}) for $N=3$:
\begin{equation}
        p(x_{1},\tau_{1}|x_{2},\tau_{2};x_{3},\tau_{3}) \; = \;
        p(x_{1},\tau_{1}|x_{2},\tau_{2}). \label{test}
\end{equation}
In figure \ref{MarkovTestI}, the contour plots of
$p(x_{1},\tau_{1}|x_{2},\tau_{2})$ and
$p(x_{1},\tau_{1}|x_{2},\tau_{2};x_{3}=0,\tau_{3})$ have been
superposed for the scales $\tau_{1}=\tau_{min}=4min$, $\tau_{2}=2
\tau_{min}$ and $\tau_{3} = 3 \tau_{min}$. The proximity of
corresponding contour lines yields evidence for the validity of
equation (\ref{test}) for the chosen set of scales. 
Additionally, two cuts through the conditional probability densities
are provided for fixed values of $x_{2}$.

%
%

\begin{figure}[H]
  \begin{center}
    \epsfig{file=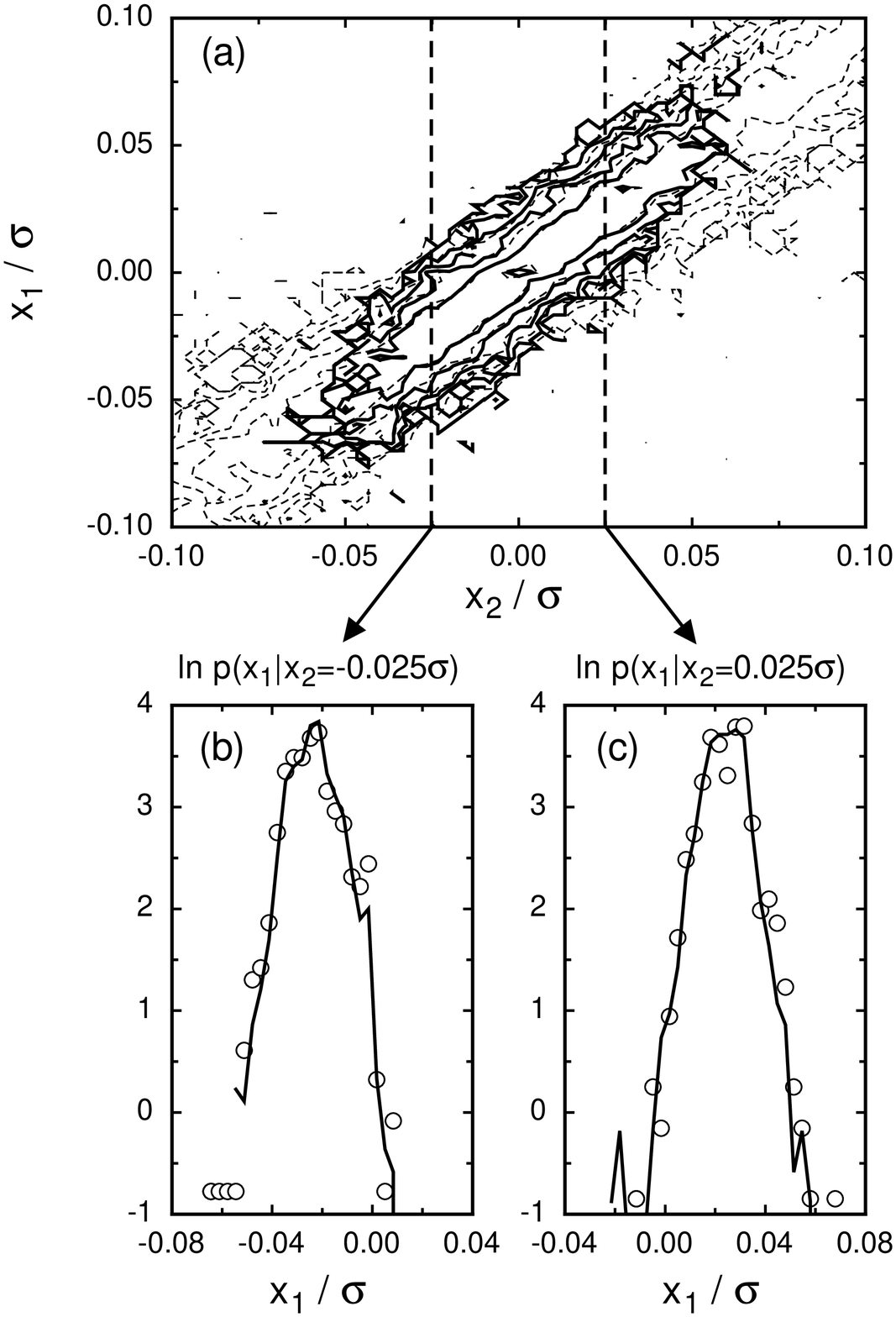, width=6.0cm}
  \end{center}
  \caption{(a): Contour plots of the conditional pdfs
  $p(x_{1},\tau_{1}|x_{2},\tau_{2})$ (dashed lines) and
  $p(x_{1},\tau_{1}|x_{2},\tau_{2};x_{3}=0,\tau_{3})$ (solid lines) for
  $\tau_{1}=1h$, $\tau_{2}=\tau_{1}+\tau_{min}$ and 
  $\tau_{3}=\tau_{1}+2\tau_{min}$.  \newline (b) and (c): Cuts
  through the conditional pdfs for $x_{2} = \pm 0.025\sigma$. Open symbols:
  $p(x_{1},\tau_{1}|x_{2},\tau_{2};x_{3}=0,\tau_{3})$, solid lines:
  $p(x_{1},\tau_{1}|x_{2},\tau_{2})$.
 }
 \label{MarkovTestII}
\end{figure}

Figure \ref{MarkovTestII} shows the same plots for a different set of 
scales: $\tau_{1}=1h$, $\tau_{2}=\tau_{1}+\tau_{min}$ and 
$\tau_{3}=\tau_{1} + 2 \tau_{min}$. Again, we find a good agreement 
between corresponding contour lines.

%
%

Similar results were obtained for several other sets of scales chosen
from the intervall $\tau_{min} \leq \tau_{1} \leq 2h$, with $\Delta
\tau = \tau_{2}-\tau_{1} = \tau_{3}-\tau_{2}$ ranging from $\tau_{min}$
to $3\tau_{min}$. 

Based on these result, we proceed with the well-founded 
assumption that the price increments of exchange rate data obey a
Markov process for the range of scales under consideration, i.e. for
scales and differences of scales larger than $\tau_{min}$.

\subsection{Kramers-Moyal Coefficients} \label{KMCoeffs}

According to equations (\ref{cpdfdef1}) and (\ref{Mndef}), the
coefficients $M_{k}(x,\tau,\Delta \tau)$ can be calculated from the
joint probability density functions. These joint pdfs
$p(\tilde{x},\tau-\Delta \tau \, ; \, x,\tau )$ are easily obtained
from the data by counting the number $N(\tilde{x}, x )$ of occurrences
of the two increments $\tilde{x}$ and $x$.

\begin{figure}[ht]
  \begin{center}
    \epsfig{file=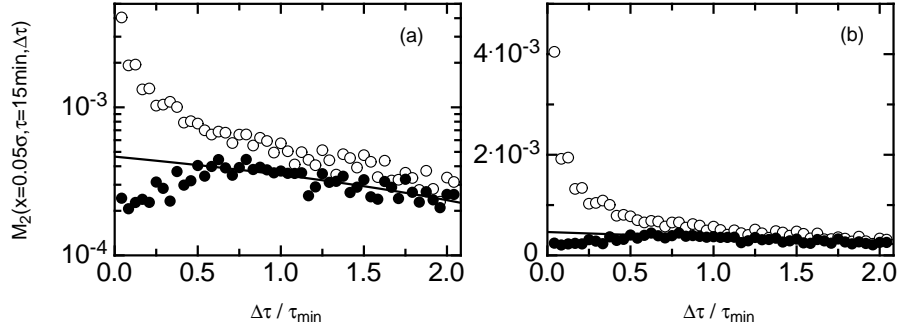, width=12.0cm}
  \end{center}
  \caption{The coefficient $M_{2}(x,\tau,\Delta \tau)$ as a function of
  $\Delta \tau$ for $x=0.05\sigma$ and $\tau = 15min$ (full circles). 
  (a): logarithmic scale, (b): linear scale.  For $\tau \geq
  \tau_{min}$, $M_{2}$ is a linear function of $\Delta \tau$ and can
  thus be extrapolated using a linear fit in the interval $\tau_{min}
  \leq \Delta \tau \leq 2 \tau_{min}$ (full line).  The open circles
  represent the results for $M_{2}$ for the same values of $x$ and
  $\tau$ when instead of the smoothed data $s(t)$ the original data
  $y(t)$ are used.  The presence of additive white noise leads to
  diverging values for $M^{(y)}_{2}$ as $\Delta \tau$ goes to zero.  
  }
 \label{ExtrapolD2}
\end{figure}

According to equation (\ref{dndef}), the limit $\Delta \tau
\rightarrow 0$ has to be performed next in order to obtain the
Kramers-Moyal coefficients. Figure \ref{ExtrapolD2} shows the
coefficient $M_{2}(x,\tau,\Delta \tau)$ for exemplarily chosen fixed
values of $x$ and $\tau$ as a function of $\Delta \tau$. Throughout
the intervall $\tau_{min} \, \leq \, \Delta \tau \, \leq 3
\tau_{min}$, the dependence of $M_{2}$ on $\Delta \tau$ turns out to be
linear. For values of $\tau < \tau_{min}/2$, the values deviate from
that linear behaviour. It is interesting to note that the range of scales 
where those deviations begin is identical with the range of scales for 
which the autocorrelation function of the reconstructed additive noise 
$n(t)$ has nonzero values (see figure \ref{AutocorrNoise}).
The limit $\Delta \tau \rightarrow 0$ is therefore performed by
fitting a straight line to the $M_{n}$ in the intervall $\tau_{min}
\leq \Delta \tau \leq 2 \tau_{min}$, thus avoiding problems with the 
values for $\Delta \tau \leq \tau_{min}/2$ (see figure
\ref{ExtrapolD2}).

Figure \ref{ExtrapolD2} also displays the coefficient $M^{(y)}_{2}$,
which is obtained when instead of the smoothed signal $s(t)$ the
original data $y(t)$ is used to calculate the coefficients $M_{k}$. 
In the presence of additive white noise, the values of $M^{(y)}_{2}$
diverge as $\Delta \tau$ goes to zero. The limit $\Delta \tau
\rightarrow 0$ can not be performed in this case. Since the
coefficients $M_{k}$ are nothing but conditional moments of the
increments (see equation (\ref{Mndef})), the reason for this behaviour
can easily be understood by expressing the second moment of
the increment of $y(t)$ in terms of $s(t)$ and $n(t)$. Using equation
(\ref{AddNoise}), we obtain:
\begin{eqnarray}
    \left< \,  \left( \, y(t+\tau) - y(t) \, \right)^2 \, \right> 
    \; & = & \; \left< \,  \left( \, s(t+\tau) + n(t+\tau) - s(t+\tau)
    - n(t+\tau) \, \right)^2 \, \right> \nonumber \\
     & = & \; \left< \,  \left( \, s(t+\tau) - s(t) \, \right)^2 \, 
     \right> \; + \; 2 \left< \, n^2 \, \right>. \label{DivProblem}
\end{eqnarray}
Due to the additive nature of the white noise $n$, the additional
constant term $2 \left< \, n^2 \, \right>$ arises which does not
depend on the scale $\tau$. Similar terms also arise for the
conditional moments $M_{k}$ when $y(t)$ is used instead of $s(t)$. 
Dividing the conditional moments of the increment by $\Delta \tau$
according to equation (\ref{Mndef}) thus leads to the diverging behaviour
of $M_{k}^{(y)}$ in the limit $\Delta \tau \rightarrow 0$ as shown in
figure \ref{ExtrapolD2}.

%
%

\begin{figure}[H]
  \begin{center}
    \epsfig{file=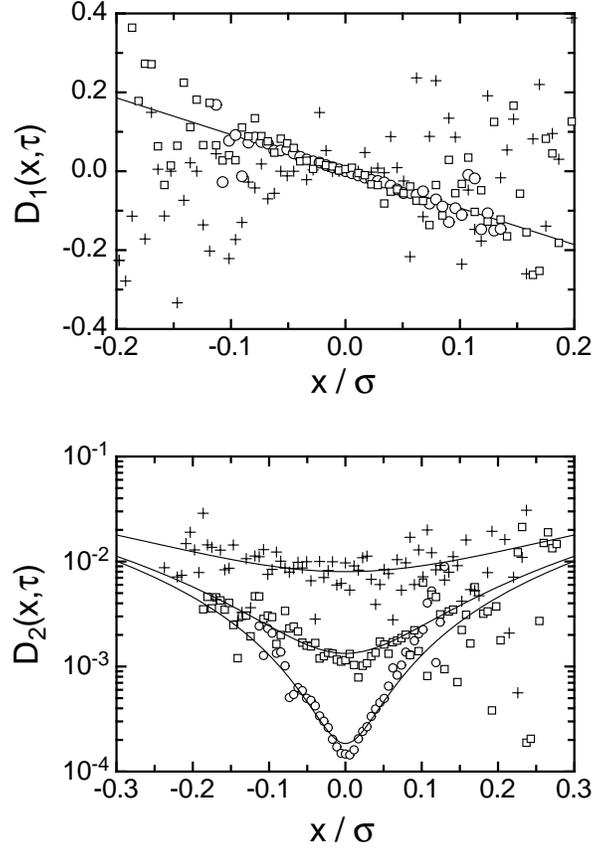, width=8.0cm}
  \end{center}
  \caption{The coefficients $D_{1}$ (a) and $D_{2}$ (b) as functions of
  the price increment $x$ at scale $\tau = 15min$ (circles), $\tau =
  2h$ (squares) and $\tau = 12h$ (crosses).
  The lines represent the results for $D_{1}$ and $D_{2}$
  obtained with the method described in section \ref{alternative} 
  (note the logarithmic presentation of $D_{2}$).
 }
 \label{DnExpUndTheo}
\end{figure}

When the smoothed signal $s(t)$ is used to calculate the coefficients
$M_{k}$, the linear dependence of $M_{k}(x, \tau, \Delta \tau)$ on
$\Delta \tau$ is found to hold for $k=1,2$, several scales $\tau$ and
all values of $x$. This enables us to calculate the coefficients
$D_{1}(x,\tau)$ and $D_{2}(x,\tau)$ from the $M_{k}$ using the method 
of linear extrapolation described above. Figure \ref{DnExpUndTheo}
shows the results for $D_{1}$ and $D_{2}$ as a function of the price
increment $x$ at several scales $\tau$. Both coefficients exhibit
simple dependencies on the price increment. While $D_{1}$ is linear
in $x$, $D_{2}$ can be approximated by a polynomial of degree two
where the linear term is zero:
\begin{eqnarray}
    D_{1}(x,\tau) & \; = \; & - \, \gamma (\tau) \, x, \nonumber \\
    D_{2}(x,\tau) & \; = \; & \alpha(\tau) \; + \; \beta(\tau) \, x^2 .
    \label{DnDirekt}
\end{eqnarray}

%
%

%
%

Equation (\ref{DnDirekt}) turns out to describe the dependencies of
the coefficients $D_{k}$ on $x$ for all scales $\tau$ up to two hours
(see figure \ref{DnExpUndTheo}). For larger scales, the statistics of
the data are too poore to allow for a proper calculation of the
coefficients $M_{k}$, which results in considerable scatter of $D_{1}$
and $D_{2}$ (as exemplarily shown for $\tau = 12h$ in figure
\ref{DnExpUndTheo}). Indeed, from the analogous analysis of turbulent
data \cite{JFM}, we know that the conditional moments $M_{k}$ can be
calculated properly only for scales $\tau$ with $\tau \leq 10^{-4} T$,
where $T$ is the total length of the data set. With $T$ being one
year for the data set considered here, we are thus restricted to
scales smaller than one hour for a direct analysis of the 
Kramers-Moyal coefficients.

Fitting the $D_{k}(x,\tau)$ by straight lines and parabolas,
respecitvely, therefore yields values for $\gamma(\tau)$,
$\alpha(\tau)$ and $\beta(\tau)$ which scatter considerably and which
do not exhibit a well defined functional dependence on the scale
$\tau$. However, what can be concluded from the analysis of the
coefficients $D_{k}$ is that the Kramers-Moyal coefficients are linear
and quadratic functions of the price increment, respectively, with
coefficients depending on the scale $\tau$.

\subsection{The Fourth Order Coefficient}

\begin{figure}[ht]
  \begin{center}
    \epsfig{file=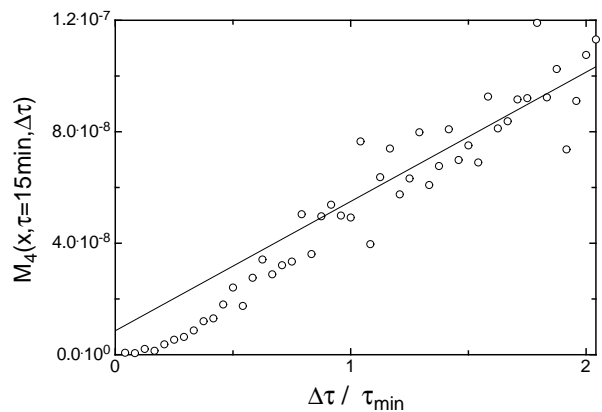, width=8.0cm}
  \end{center}
  \caption{The coefficient $M_{4}(x,\tau,\Delta \tau)$ as a function of
  $\Delta \tau$ for $x = -0.05 \sigma$ and $\tau = 15min$
  (circles). 
  }
 \label{ExtrapolD4}
\end{figure}

According to Pawula's theorem, it is of importance to estimate the
fourth order coefficient and to decide whether it may be neglected. 
Figure \ref{ExtrapolD4} shows the coefficient $M_{4}(x, \tau, \Delta
\tau)$ for fixed values of $x$ and $\tau$ as a function of $\Delta
\tau$.  Again, we find a linear dependence of the coefficient on
$\Delta \tau$ for $\Delta \tau \geq \tau_{min}$.  But whereas
$M_{2}(\Delta \tau)$ increases as $\Delta \tau$ goes to zero, $M_{4}$
{\it decreases}.  Furthermore, the linear extrapolation yields a value
for $D_{4}$ which is small compared to the values of $M_{4}$ for
$\tau_{min} \leq \tau \leq 2 \tau_{min}$.  This is a first hint that $M_{4}$
tends to zero in the limit $\Delta \tau \rightarrow 0$.

%
%

\begin{figure}[ht]
  \begin{center}
    \epsfig{file=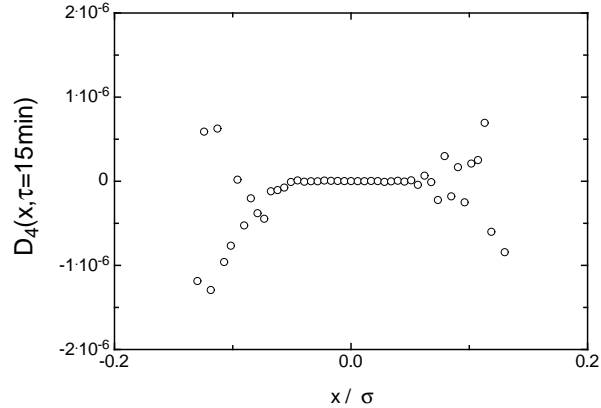, width=8.0cm}
  \end{center}
  \caption{The fourth coefficient $D_{4}(x,\tau)$ as a function of
  $x$ for scale $\tau = 15min$ (circles). The linear extrapolation 
  yields positive as well as negative values for $D_{4}$.
  }
 \label{D4Result}
\end{figure}

Performing the limit $\Delta \tau \rightarrow 0$ for various values of $x$ 
using the linear extrapolation confirms the above results. We obtain 
positive as well as negative values for $D_{4}$, as shown in figure 
\ref{D4Result} for the scale $\tau = 15min$. Since, by definition, $D_{4}$ 
is a positive quantity, this means that the data set allows to take 
$D_{4}$ to be zero.

%
%

Similar results were obtained for several scales $\tau \leq 2h$. Based on 
these results, we will proceed with the assumption that the fourth 
order coefficient $D_{4}$ vanishes, i.e. that Pawula's theorem 
applies and that the stochastic process is described by the Fokker-Planck 
equation.

\subsection{An Alternative Approach} \label{alternative}

Based on the results of the preceeding section, we will proceed with
the assumption that the statistics of the price increment is governed
by the Fokker-Planck euqation with coefficients $D_{1}$ and $D_{2}$
given by equation (\ref{DnDirekt}) and try to find estimates for the
coefficients $\gamma$, $\alpha$ and $\beta$.

For the slope $\gamma(\tau)$ of $D_{1}$, such an estimate can be
obtained from the conditional first order moment $T^{1}(\tau) = 
\left< \, x(\tau) \, | x_{0}(\tau_{0}) \, \right>$ of the price increment. 
Multiplying the Fokker-Planck equation (\ref{foplacond}) with $x$, 
integrating with respect to $x$ and finally using the result 
(\ref{DnDirekt}) for $D_{1}$, an equation for $T^{1}(\tau)$ 
can be derived. The result is:
\begin{eqnarray}
    - \tau \frac{\partial}{\partial \tau} \, T^{1}(\tau) \; & = & 
    - \, \gamma(\tau) \, T^{1}(\tau) . \label{CondExpEqu}
\end{eqnarray}
In deriving equation (\ref{CondExpEqu}), it was assumed that $D_{2}(x,\tau) 
\, p(x,\tau | x_{0},\tau_{0})$ goes to zero for $|x| \rightarrow \infty$.

With the initial condition $T^{1}(\tau = \tau_{0}) = \left< x(\tau
= \tau_{0}) | x_{0}(\tau_{0}) \right> = x_{0}$, equation
(\ref{CondExpEqu}) is solved by:
\begin{eqnarray}
    T^{1}(\tau) \; = \; x_{0} \, \exp \left\{ \,
    \int\limits_{\tau_{0}}^{\tau} \,
    \frac{\gamma(\tau')}{\tau'} \, d\tau' \right\}. 
    \label{CondExpSol}   
\end{eqnarray}
If the conditional moment $T^1$ follows the simple prediction
(\ref{CondExpSol}), this can be taken as a strong hint for the
validity of the assumptions underlying the derivation (validity of the
Fokker-Planck equation and linear dependence of $D_{1}$ on $x$). On
the other hand, the simple dependence of $T^{1}(\tau)$ on $\gamma$
could also be used to measure the function $\gamma(\tau)$.

\begin{figure}[ht]
  \begin{center}
    \epsfig{file=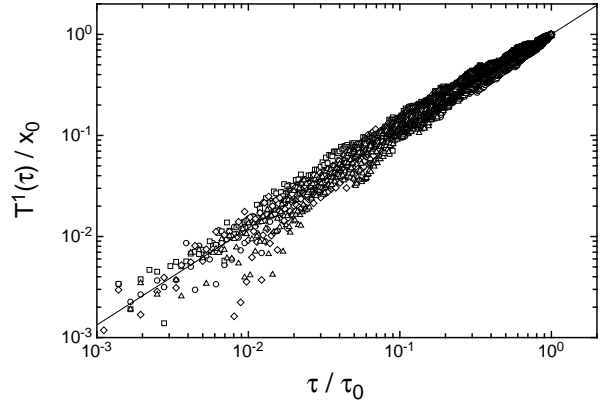, width=8.0cm}
  \end{center}
  \caption{The conditional moment $T^1$ (divided by $x_{0}$) as a
  function of the time scale $\tau$ for $\tau_{0}=3h$, $x_{0} =
  -0.05\sigma$ (circles), $x_{0} = -0.025\sigma$ (squares), $x_{0} =
  0.05\sigma$ (diamonds) and $x_{0} = 0.05\sigma$ (triangles). The
  straight line represents a power law fit to the data yielding a
  scaling exponent of $0.95$.
 }
 \label{BedingteMomente}
\end{figure}

Figure \ref{BedingteMomente} shows the conditional expectation 
values $T^{1} / x_{0}$ as a function of $\tau / \tau_{0}$ for scale 
$\tau_{0} = 3h$ and several values of $x_{0}$. Indeed, as was expected 
from equation (\ref{CondExpSol}), $T^1$ divided by $x_{0}$ turns 
out to be an universal function of $\tau / \tau_{0}$. In particular, we 
do not find a systematic dependence on $x_{0}$.

%
%

Furthermore, it is found that $T^1$ is well described by a power law 
in $\tau$, i.e.: $T^1 \propto \tau^{c}$. It is easily seen that such a 
power law arises from equation (\ref{CondExpSol}) for the case 
that $\gamma(\tau)$ is constant:
\begin{eqnarray}
    T^{1}(\tau) \; & = & \; x_{0} \, \exp \left\{ \,
    \int\limits_{\tau_{0}}^{\tau} \,
    \frac{\gamma}{\tau'} \, d\tau' \right\} \; = \; x_{0}  \, \exp \left\{ 
    \, \gamma \ln\left( \frac{\tau}{\tau_{0}} \right) \, \right\} 
    \nonumber \\ & = & \; x_{0} \, \left( \frac{\tau}{\tau_{0}} 
    \right)^{\gamma}. \label{T1Sol}
\end{eqnarray}
Hence, the conditional first order moment $T^1$ of the price increment
$x$ provides a very convenient method to determine $\gamma$. Fitting
the data presented in figure \ref{BedingteMomente}, we obtain a value
of $\gamma = 0.95$ for the scaling exponent. Summarizing the results
obtained by varying the values of $x_{0}$ and $\tau_{0}$ finally yields:
\begin{eqnarray}
    \gamma \; = \; 0.93 \pm 0.02 \; . \label{gamma}
\end{eqnarray}
It is interesting to note that the value of $\gamma$ we obtain is
close to one. However, the statistics of the data set we used is
rather poor and further investigations using larger data sets will be
necessary to decide whether $\gamma$ deviates significantly from one.

Estimates for the coefficients $\alpha(\tau)$ and $\beta(\tau)$ of
$D_{2}$ can be obtained using the equations for the n-th order moments
$<x(\tau)^n>$ of the price increment. With the result (\ref{DnDirekt}) 
for $D_{1}$ and $D_{2}$, equation (\ref{momentengleichung}) yields for 
$n\geq2$:
\begin{eqnarray}
    - \tau \frac{\partial}{\partial \tau} \left< x^n(\tau) \right> 
    & \; = \; & - \, n \, \gamma \, \left< x^n(\tau) \right> \; + \; n \, 
    (n-1) \, \alpha(\tau) \, \left< x^{n-2}(\tau) \right> \nonumber \\
	&& \; + \,   n \, (n-1) \, \beta(\tau) \, \left< x^n(\tau) \right>, 
\end{eqnarray}	
which can be rewritten as:
\begin{eqnarray}       
	\frac{\tau \frac{\partial}{\partial \tau} \left< x^n(\tau) \right> }{n
	\, \left< x^n(\tau) \right> } & \; = \; & \gamma \; - \; (n-1) \,
	\beta(\tau) \; - \; (n-1) \, \alpha(\tau) \, \frac{\left< x^n(\tau)
	\right>}{\left< x^{n-2}(\tau) \right>}.  \label{StrucFuncs1}
\end{eqnarray}
As $\gamma$ is known from the analysis of the conditional first order 
moment, two of the equations (\ref{StrucFuncs1}) are sufficient to 
calculate the two unknown functions $\alpha(\tau)$ and $\beta(\tau)$ from 
the moments $\left< x^n(\tau) \right>$. We use the moments of order two 
and four, since even moments of low order can be determined best from 
empirical data. Due to the rather limited number of samples it is even 
uncertain whether the moment of order six can be determined properly (see 
\cite{JFM} for a more detailed discussion of this topic).

\begin{figure}[ht]
  \begin{center}
    \epsfig{file=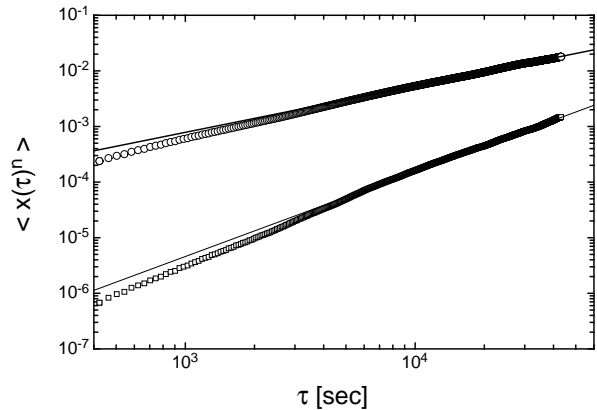, width=8.0cm}
  \end{center}
  \caption{The moments $\left< x^n(\tau) \right>$ of the price increment 
  $x(\tau)$ as functions of the time scale $\tau$ for order $n=2$ (top) 
  and $n=4$ (bottom). Throughout the range $\tau_{min} \; \leq \; \tau \; 
  \leq 12h$ the moments are in good approximation described by power laws 
  (straight lines).
 }
 \label{StrucFuncScaling}
\end{figure}

Figure (\ref{StrucFuncScaling}) shows the moments $\left< x^n(\tau) \right>$ 
for order two and four as functions of the time scale $\tau$. Throughout the 
range $\tau_{min} \; \leq \; \tau \; \leq 12h$, the moments can be 
described by power laws in $\tau$:
\begin{eqnarray}
	\left< x^n(\tau) \right> \; \propto \; \tau^{\zeta_{n}}.
	\label{Scaling}
\end{eqnarray}
Using this result, the left hand side of equation (\ref{StrucFuncs1}) 
can easily be shown to be equal to $\frac{\zeta_{n}}{n}$. 
Fitting the empirically determined moments with power laws, we find the 
following values for the scaling exponents $\zeta_{n}$:
\begin{eqnarray}
	\zeta_{2} \; & = & \; 0.94 \, \pm \, 0.04 \nonumber \\
	\zeta_{4} \; & = & \; 1.72 \, \pm \, 0.12. \label{Exponents}
\end{eqnarray}
The estimates for the errors were obtained by varying the range of
scales used for the fit. Note that we found a value of $\zeta_{2}$
which is slightly smaller than one, the value which is usually assumed
\cite{nature96, ReplyStanley}. Due to the poor statistics of the data set,
however, this result is by no means significant; the second order
moment can be fitted with a scaling exponent of $\zeta_{2} =1$ with
almost the same accuracy. Yet, the values given in (\ref{Exponents})
are the best fitting coefficients for the data set under consideration
and will therefore be used here.

%
%

Using the results for the second and fourth order moments, equations 
(\ref{StrucFuncs1}) can easily be solved and yield estimates for the 
unkown functions $\alpha(\tau)$ and $\beta(\tau)$. We find that $\alpha$ 
is a linear function of the time scale $\tau$ with a slope $\alpha_{0}$ 
of $0.016 d^{-1}$, while $\beta$ is approximately constant in $\tau$ and 
has a value of $0.11$. 
Let us summarize our results for $D_{1}$ and $D_{2}$:
\begin{eqnarray}
	D_{1}(x,\tau) \; & = & \; - \, \gamma \, x \nonumber \\
	\gamma \; & = & \; 0.93 \, \pm\, 0.02 \nonumber \\
	D_{2}(x,\tau) \; & = & \; \alpha_{0} \, \tau \; + \; \beta \, x^2
	\nonumber \\
	\alpha_{0} \; & = & \; 0.016 d^{-1} \, \pm \, 0.002 d^{-1}
	\nonumber \\
	\beta(\tau) \; & \approx & \; const \; = \; 0.11 \, \pm \, 0.02. 
	\label{D12Coeffs}
\end{eqnarray}

\subsection{Consistency Checks}

The results and assumptions of the preceeding sections lead to Fokker-
Planck equations for the pdf $p(x,\tau)$ and the conditional pdf 
$p(x,\tau|x_{0},\tau_{0})$, respectively (equations (\ref{foplauncond}) and 
(\ref{foplacond})). The coefficients $D_{1}$ and $D_{2}$ which completely 
determine these equations were estimated from the data. Thus we can 
claim to have found the complete description of the stochastic process 
for the data.
In order to test this result, we compare the (numerical) solution of
the Fokker-Planck equation with the distributions obtained directly
from the data.  The algorithm used for the numerical iteration is
based on the approximative solution of the Fokker-Planck equation for
small steps $\Delta \tau$.  According to \cite{Risken}, the
conditional pdf $p(x_{1},\tau_{0}-\Delta \tau|x_{0},\tau_{0})$ is, for
small $\Delta \tau$ and arbitrary $D_{1}$ and $D_{2}$, a Gaussian
distribution in $x_{1}$ with mean value $x_{0}-D_{1}\Delta \tau$ and
standard deviation $\sqrt{2D_{2}\Delta \tau}$.  In order to obtain the
conditional densities for larger steps, we use the Chapman-Kolmogorov
equation
\begin{equation}
	p(x_{2},\tau_{0}-2\Delta \tau | x_{0},\tau_{0})  =  
	\int\limits_{-\infty}^{+\infty}  p(x_{2},\tau_{0}-2\Delta \tau | 
	x_{1},\tau_{0}-\Delta \tau)  p( x_{1},\tau_{0}-\Delta \tau | 
	x_{0},\tau_{0}) dx_{1} . \label{ChapKol}
\end{equation}
(\ref{ChapKol}) is a direct consequence of the Markov condition (\ref{markovcondtheo}). 
Iterating this procedure, we finally obtain the conditional pdf 
$p(x,\tau_{0}-n\Delta \tau | x_{0}, \tau_{0})$. Multiplying with 
$p(x_{0},\tau_{0})$ and integrating with respect to $x_{0}$ yields the 
pdf $p(x,\tau_{0}-n \Delta \tau)$. 

\begin{figure}[H]
  \begin{center}
    \epsfig{file=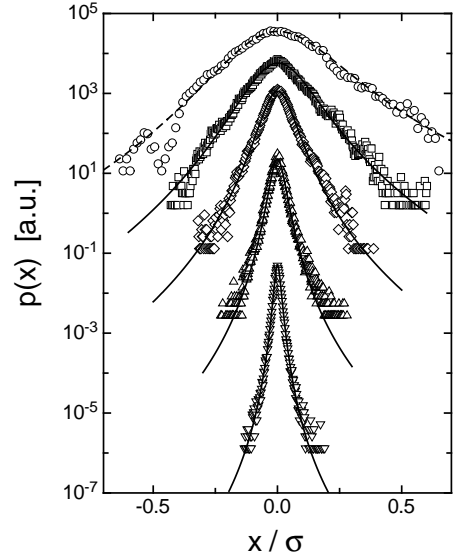, width=6.0cm}
  \end{center}
  \caption{Comparison of the numerical solutions of the Fokker-Planck 
  equation (solid lines) for the pdfs $p(x,\tau)$ with the pdfs 
  obtained directly from the data (open symbols). The scales $\tau$ 
  are (from top to bottom): $\tau \; = \; 12h, \, 4h, \, 1h, \, 15min$ 
  and $4min$. The pdf at the largest scale $\tau = 12h$ was 
  parametrized (dashed line) and used as inital condition for the 
  iteration of the Fokker-Planck equation. Curves are shifted in 
  vertical direction for clarity of presentation.
  }
 \label{TheoExp1}
\end{figure}

Figure \ref{TheoExp1}, which compares the solutions of the Fokker-Planck 
equation for the pdf $p(x,\tau)$ with the empirically estimated pdfs, 
proves that the Fokker-Planck equation accurately describes the evolution 
of $p(x,\tau)$ in $\tau$ over the range $4min \leq \tau \leq 12h$.

%
%

\begin{figure}[H]
  \begin{center}
    \epsfig{file=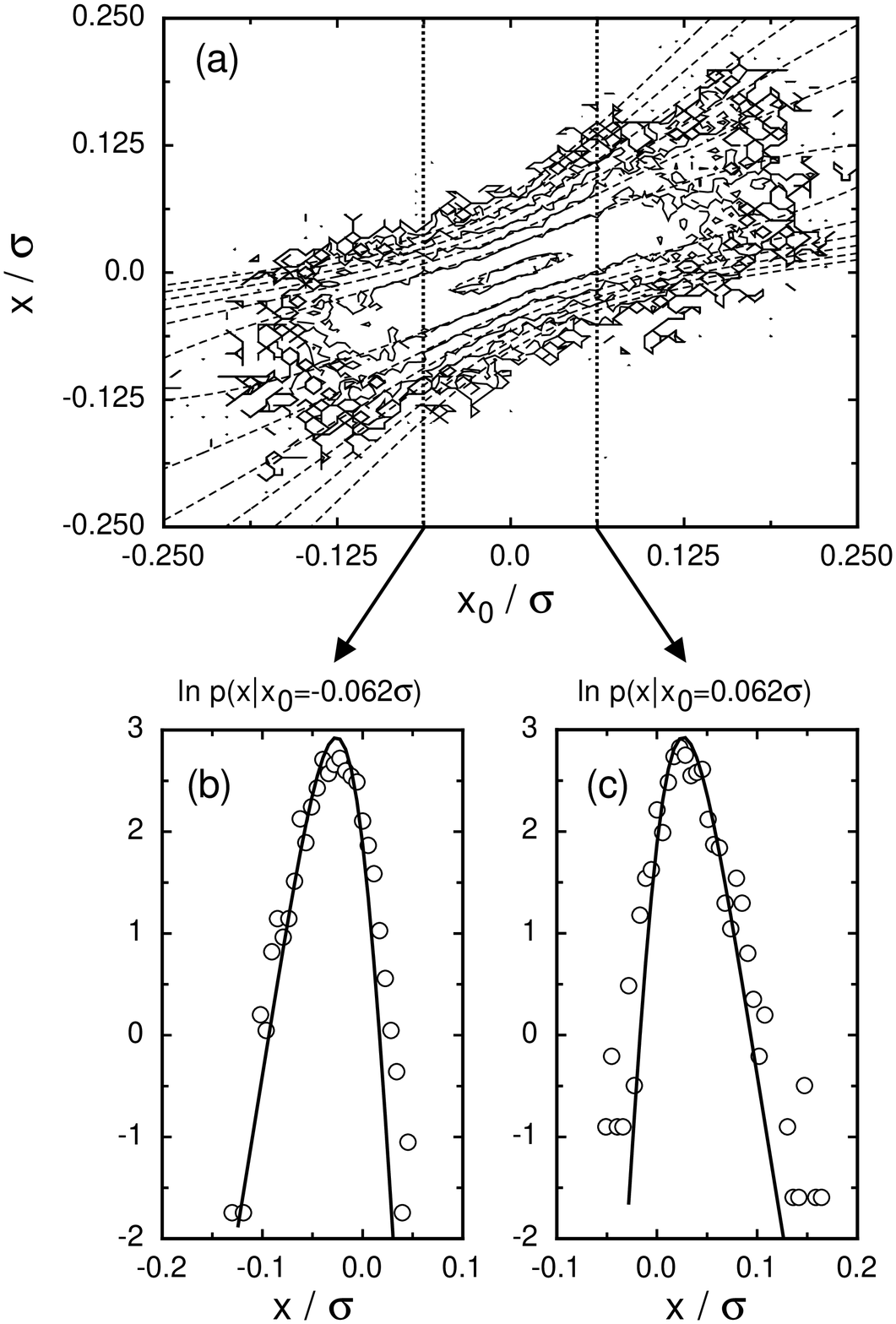, width=8.0cm}
  \end{center}
  \caption{Comparison of the numerical solution of the Fokker-Planck 
  equation for the conditional pdf $p(x,\tau|x_{0},\tau_{0})$ with 
  the emprical data. \newline
  (a): Contour plots of $p(x,\tau|x_{0},\tau_{0})$ for $\tau_{0}=1h$ 
  and $\tau=0.5h$. Dashed lines: numerical solution of eq. 
  (\ref{foplacond}), solid lines: empricial data. \newline
  (b) and (c): Cuts through $p(x,\tau|x_{0},\tau_{0})$ for 
  $x_{0}=-0.062 \sigma$ and $x_{0}=+0.062 \sigma$, respectively. Open 
  symbols: empirical data, solid lines: numerical solution of the 
  Fokker-Planck equation.
  }
 \label{TheoExp2}
\end{figure}

As mentioned above, the Fokker-Planck equation also governs the 
conditional pdf $p(x,\tau|x_{0}, \tau_{0})$. As a further test of our 
results, we calculated the solutions of the Fokker-Planck equation 
(\ref{foplacond}). Figure \ref{TheoExp2} shows the result for 
$\tau_{0}=1h$ and $\tau=0.5h$, again in comparison with empirical 
data. Taking into account the various uncertainties and assumptions in the 
determination of the coefficients $D_{1}$ and $D_{2}$, the agreement 
between the solution of (\ref{foplacond}) and the data is remarkably good.

%
%

\section{Summary and Conclusions} \label{discussion}

The purpose of the present article is to show how the mathematical 
framework of Markov processes can be applied to the analysis of emprical 
high frequency exchange rate data. Within the limitations due to the 
finite number of samples, we were able to verify the Markovian properties 
of the price increment for time delays $\tau$ larger than $\tau_{min} = 
4min$.

Furthermore, we obtained estimates for the Kramers-Moyal coefficients 
$D_{k}(x,\tau)$ and found hints that the fourth order coefficient 
$D_{4}$ is zero. The comparison of the solutions of the resulting Fokker-
Planck equations with the empricial distributions strongly supports our 
results for $D_{1}$ and $D_{2}$.

It is worth to note that the results given in the present article differ 
from the results of our previous analysis presented in \cite{PRL}. While 
in \cite{PRL} $D_{1}$ and $D_{2}$ were also found to be linear and quadratic 
functions of $x$, respectively, the values given for the coefficients 
$\gamma$, $\alpha$ and $\beta$ were different from the ones given here in 
equation (\ref{D12Coeffs}). 

The analysis presented in \cite{PRL} is incomplete, because it did not take 
into account the presence of the additive white noise (see section 
\ref{processing}) and mainly focused on the functional dependence of the 
$D_{k}$ on $x$. The numerical values of the coefficients $\gamma$, $\alpha$ 
and $\beta$ were chosen such that the pdfs of the price increment $x$ were 
fitted best by the resulting Fokker-Planck equation. However, it can be 
shown that the coefficients given in \cite{PRL} fail to describe the 
conditional pdf $p(x,\tau|x_{0},\tau_{0})$, i.e. cannot be correct. 

This is a quite important result: There are several choices for the 
coefficients which correctly describe the one-point distribution 
$p(x,\tau)$, but only one choice which also describes the joint statistical 
properties of several increments on different scales. This, in turn, means 
that a complete statistical characterization of the price increments (or 
returns) necessarily has to include the multi-scale statistics and must not 
be restricted to the analysis of increments (or returns) on only one 
scale. 

The results of our analysis presented in this paper also eliminate the 
discrepancy between our approach as presented in \cite{PRL} and the often 
reported power-law behaviour of the distributions of returns. As previously 
shown in \cite{Sornette}, the solutions of the Fokker-Planck equation 
asymptotically show scaling behaviour with a scaling exponent $\mu$ which 
can be calculated from the linear and quadratic coefficient of $D_{1}$ and 
$D_{2}$, respectively:
\begin{eqnarray}
	\mu \; = \; \frac{\gamma}{2 \, \beta}. \label{MuEquation}
\end{eqnarray}
While the values reported in \cite{PRL} lead to a value of about $12$ for 
the scaling exponent $\mu$, in disagreement with the usually observed value 
between 3 and 5, our new values (\ref{D12Coeffs}) yield $\mu \, = \, 
4.2 \pm 0.8$. 

Another interesting result of our analysis concerns the symmetry of the 
stochastic process underlying the evolution of the price increment $x$ in 
the scale variable $\tau$. According to (\ref{D12Coeffs}), the drift 
coefficient $D_{1}$ is completely antisymmetric in $x$ ($D_{1}(-x) \, = \,
- D_{1}(x)$), while the diffusion coefficient $D_{2}$ is symmetric in 
$x$. The resulting Fokker-Planck 
equation (\ref{foplauncond}) for the pdf $p(x,\tau)$ is symmetric in 
$x$. This means that if the large scale pdf, which serves as the initial 
condition for the partial differential equation (\ref{foplauncond}), was 
symmetric, it would remain symmetric for all scales $\tau$. In other words: 
The asymmetries of the distributions of price increments on small scales are 
solely due to the asymmetry of the increment at the largest scale, i.e. they 
are a consequence of some market event on long terms. The stochastic process itself, at 
least on scales smaller than a day, is symmetric in $x$. This result might 
provide a possibility to distinct between trends (as for example caused 
by the long term development of a country's economy) and the short term 
fluctuations caused by the market itself i.e. by the interactions of 
several agents trading an asset. This fits well with the results of commonly 
discussed market models, which are quite succesfull in qualitatively 
explaining most of the features of real markets but which do not naturally 
exhibit asymmetries between positive and negative returns, see for example 
\cite{ContBouchaud} and \cite{StaufferSornette}. Our result indicates that 
there is no need to incorporate asymmetries in such models. However, so far 
we can only present results for a single data set over an one-year period. 
Further studies on a larger variety of assets and time scales will be 
necessary to check the significance of this particular result. 

Finally, we would like to comment on the proposed analogy between financial 
markets and turbulent hydrodynamic flows, cf. \cite{nature96}. 
The complex statistical behaviour of velocity increments on a certain length 
scale in turbulent flows is assumed to be due to a cascading process. Energy, 
which is fed into the system on large scales, is continuously transported 
towards smaller scales due to the inherent instability of vortices of a given 
scale towards perturbations on smaller scales. Finally, the energy is 
dissipated at the smallest scale. 

A similar mechanism has been proposed for financial markets, where the 
energy cascade was replaced by a flow of information. Initially, the 
assumption of a cascading process in financial markets was based on 
similarities in the empirical description of the pdfs of price and 
velocity increments, respectively \cite{nature96}. The analogy between 
turbulence and finance has inspired many further studies (see, for example, 
\cite{ArneodoFinanz}, \cite{SchertzerFinanz}, \cite{BreymannKaskade}), but 
has also been criticized for being too superficial \cite{Stanley}. 

However, the results of the Markov analysis of financial and turbulent data 
(see \cite{JFM} for details on the Markov analysis of turbulence) allow for 
a more detailed comparison of the stochastic processes underlying the two 
phenomena with the predictions of common cascade models. 

The most prominent model for turbulence is Kolmogorov's theory dating back 
to 1962 (furtheron refered to as K62). The prediction of the K62 model for 
the coefficients $D_{1}$ and $D_{2}$ can easily be shown to be \cite{PRL1}:
\begin{eqnarray}
	D_{1} & \; = \; & - \, \gamma \, x, \nonumber \\
	D_{2} & \; = \; & \beta \, x^{2}, \label{K62}
\end{eqnarray}
where $\gamma$ and $\beta$ both are constant ($\gamma \approx 1/3$ for 
turbulence).
Castaing \cite{CastaingKaskade} proposed the model of a multiplicative 
cascade where the increment on a given scale is linked to an increment 
at a larger scale via a random multiplier. This model was shown to 
imply Markovian properties \cite{Amblard} and to predict coefficients $D_{1}$ 
and $D_{2}$ similar to those predicted by K62, except that it allows for a 
scale dependence of $\gamma$ and $\beta$.

Recently, Dubrulle \cite{Dubrulle} proposed an extension of Castaing's 
multiplicative cascade model which incorporates additive noise. 
Comparing the results (\ref{D12Coeffs}) of our analysis with the predictions 
of the various models, we can state that the statistics of the price 
increment is well described by the model of a multiplicative cascade with 
additive noise (or, in the notion of \cite{Sornette}, by the mechanism of 
"multiplicative noise with reinjection").

It is interesting to note the result of the Markov analysis for fully
developed turbulent flows \cite{JFM}. It turns out that measured
turbulent data are not completely described by any of the above
mentioned models. While $D_{1}$ turns out to be linear, 
$D_{2}$ exhibits are more complicated
dependence on the increment than predicted by the multiplicative
cascade models. In particular, we find a linear term in $D_{2}$ which
breaks the symmetry and which is by no means negligable, see
\cite{JFM}. In some sense, one might therefore say that financial
data are much closer to ideal "Kolmogorov-like" turbulence than turbulent data.
Besides this analogy we found a clear differene in the mechanisms 
leading to the finite step-size of the cascade processes. In the case 
of turbulence a smoothing out due to viscosty was found, while 
financial markets seem to be dominated by some uncoorelated 
additional noise on the smallest scales.

Summarizing, it is the concept of a cascade in time hierarchy that
allowed us to derive the results of the present paper, which in turn
quantitatively supports the initial concept of an analogy between
turbulence and financial data.  Furthermore, we have shown that the
smooth evolution of the pdfs down along the cascade towards smaller
time delays is caused by a Markov process with multiplicative noise.

Metaphorically, our characterization of volatility of currency markets
as random and uneven flow of information to the markets is similar to
\cite{Gallant} which describes stochastic process
of the stochastic volatility model as random and uneven flow of news
to the financial markets.

\section* {Acknowledgements}
We gratefully acknowledge useful discussions with A. Soofi, D. Sornette, Zhi-Feng
Huang, St. L{\"u}ck and M. Siefert. The FX data set has been provided by {\em Olsen \&Associates} 
(Z\"urich).

\end{document}